\documentclass [prl,twocolumn,showpacs,superscriptaddress]{revtex4}
\usepackage {graphicx}
\usepackage {dcolumn}
\usepackage {bm}

\begin {document}
\title {First principle computation of stripes in cuprates}
\author {V.I. Anisimov}
\affiliation {Institute of Metal Physics, 620219 Ekaterinburg GSP-170, Russia}
\author {M.A. Korotin}
\affiliation {Institute of Metal Physics, 620219 Ekaterinburg GSP-170, Russia}
\author {A.S. Mylnikova}
\affiliation {Institute of Metal Physics, 620219 Ekaterinburg GSP-170, Russia}
\affiliation {Ural State Technical University, 620002 Ekaterinburg, Russia}
\author {A.V. Kozhevnikov}
\affiliation {Institute of Metal Physics, 620219 Ekaterinburg GSP-170, Russia}
\affiliation {Ural State Technical University, 620002 Ekaterinburg, Russia}
\author {J. Lorenzana}
\affiliation { INFM Center for Statistical Mechanics and Complexity,
Universit\`a di Roma ``La Sapienza'', P. Aldo Moro 2, 00185 Roma, Italy}
\date {\today}

\begin {abstract}
We present a first principle computation of vertical stripes in
La$_{15/8}$Sr$_{1/8}$CuO$_4$ within the LDA+$U$ method. We find
that Cu centered  stripes are unstable toward O centered stripes.
The metallic core of the stripe is quite wide and shows reduced
magnetic moments with suppressed antiferromagnetic (AF)
interactions. The system can be pictured as alternating metallic
and AF two-leg ladders the latter with strong AF interaction and a
large spin gap. The Fermi surface shows warping due to interstripe
hybridization. The periodicity and amplitude of the warping is in
good agreement with angle resolved photoemission experiment. We
discuss the connection with low-energy theories of the cuprates.
\end {abstract}


\maketitle

Recent theories of the high temperature superconducting cuprates are based
on the idea that hole-rich quasi one-dimensional textures (stripes) are the
basic building blocs to understand the low-energy physics\cite {cas95b,
zaa96b, eme97, gra99, eme00, liu01, kro97, kiv98, voj99, sac03, arr03,
lor02b, lor03}. These model computations either assume stripes to
obtain a low-energy quantum field theory (QFT)\cite {eme97, gra99, eme00,
liu01, kro97, kiv98, net01, arr03} or obtain stripes in simplified model
Hamiltonians\cite {cas95b, zaa96b, voj99, sac03, lor02b, lor03}. For the
QFT's detailed information about the symmetry and the extent of real
stripes is lacking so much of the modeling is based on guesses and the
relevant regimes are identified a posteriori, by confrontation with
experiment\cite {tra95, tra97, yam98, ara99, ara00}, leading to a plethora
of possible scenarios.

In order to constrain low-energy models some progress can be made by
solving more or less realistic  Hamiltonians in various approximations\cite
{zaa89, poi89, mac89, hsch90, zaa96, cas95b, voj99, lor02b, lor03} however
these are tight to strong assumptions about the electronic structure since
are restricted to a small number of orbitals and/or short range
interactions and a single Cu-O plane. Here we present a first principle
computation of stripes in cuprates based on the LDA+$U$ method for doping
$x=1/8$ and the experimentally measured magnetic incomensurability
$\epsilon=1/8$. Our computation are based on the full orbital variational
space and the three-dimensional Coulomb interaction and so go beyond the
previous works. LDA based methods are widely believed to give reliable
charge distributions and therefore can provide a realistic picture of
stripe textures. We find that a Cu centered solution is not stable. Instead
the stripe is O centered and consist of two columns of ``metallic'' Cu (the
core) and two columns of antiferromagnetic (AF) Cu forming an alternating
system of metallic and AF two-leg ladders.

Added holes are spread over the ``metallic'' Cu rows and the surrounding O
leading to a stripe core much wider than the traditional stripe
picture\cite {tra97}. The electronic structure consists of a half-filed
quasi one-dimensional band for each stripe core corresponding to motion
along the stripe. The system can be schematized as an array of
one-dimensional electron gases (1DEG) separated by spin-ladders (``the
environment'' in the language of Ref.~\cite {eme97,kiv98,gra99,liu01}) with a
large spin-gap. Interstripe hybridization is not negligible but leads to a
noticeable warping of the Fermi surface which is in good agreement with
angle resolved photoemission experiments\cite {zho99}. Additionally there
is approximate particle-hole symmetry around the Fermi level in good
agreement with transport experiments\cite {nod99, wan01}. We provide 
 estimates of the relevant energy scales for ``striped''
low-energy theories of cuprates.

\begin {figure}[htbp]
\includegraphics [width=\linewidth]{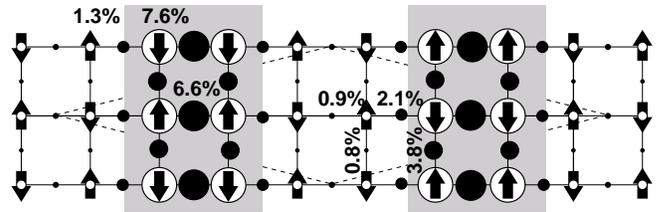}
\caption {The hole-charge and spin order within CuO$_2$ plane. Sites with
(without) arrows (indicating spin directions) are Cu (O). The radii of
circles are proportional to the hole content inside atomic spheres. The
grey regions are  metallic core with Cu of type Cu1 and the white regions
are the AF ladders with Cu of type Cu2. Numbers indicate the distribution
of the doped hole. Not shown are 1,2~\% (0,6~\%) of hole charge in the
apical O's belonging to Cu1 (Cu2) and 0,3~\% of charge on La. The unit cell
in the CuO$_2$ plane is shown by the dashed line.}
\label {fig:rho}
\end {figure}

In the present work we used a  LDA+$U$ approach\cite{ani91,ani97}
realized in frame of scalar relativistic LMTO method within the Atomic
Sphere Approximation (ASA)\cite {and75} with the Coulomb term of LDA+$U$
functional proposed in Ref.~\cite {ani91}. The radii of atomic spheres were
r(Cu)=2.2~a.u., r(O-plain)=1.94~a.u., r(O-apex)=2.00~a.u., r(La)=3.2~a.u;
three types of empty spheres were added. The values of on-cite Coulomb $U$
and exchange $J$ parameters were chosen 8 and 0.88~eV, correspondingly.

We studied a supercell consisted of eight formula units with an
additional hole in the absence of the Sr impurity, compensating
the charge by adding suitable negative charges to all atoms in the
supercell.
A Cu centered solution could not be stabilized, not even as a metastable
state. In Fig.~\ref {fig:rho} we show the charge and spin distribution of
the stable oxygen centered solution. The hole charge is concentrated on the
two core Cu's forming the AF domain wall and the surrounding oxygens (grey
region) and leak significantly over the AF regions indicating that
interstripe overlap is not negligible as discussed below.

\begin {figure}
\includegraphics [width=9cm,angle=0,clip=true]{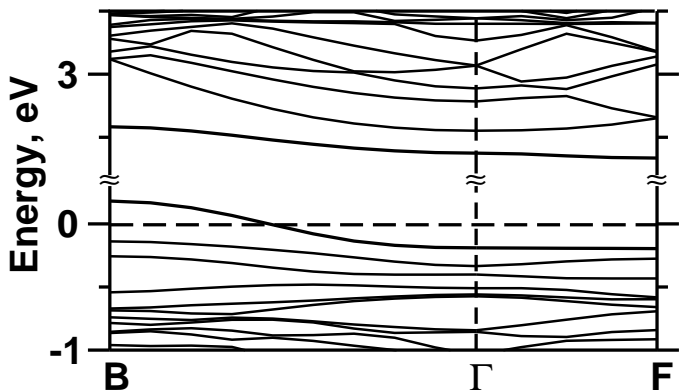}
\caption {Calculated band structure of La$_{15/8}$Sr$_{1/8}$CuO$_4$
around the Fermi level (E$_F$=0). Left (right) panel corresponds to  the
direction parallel (perpendicular) to the stripe. The point B
corresponds to momentum $[0,\pi/(2a)]$ in the Cu-O plane.
}
\label {bands}
\end {figure}

The stripe core consist of rungs of ferromagnetic Cu sites. This
locally enhances the effective Cu-Cu hybridization and splits
states into the charge transfer band which disperse strongly in
the direction of the stripe (Fig.~\ref{bands}). The band crossing
the Fermi level (hereafter ``the active band''), is half-filled,
has quasi one-dimensional character and is similar to the one
found in the three-band Hubbard model\cite {lor02b}. The
electronic structure is approximately particle-hole symmetric
around the chemical potential. This is in agreement with the
proposal  that approximate particle-hole symmetry in the stripe
state is the reason why Hall effect\cite {wan01, nod99} and the
thermopower\cite {nod99} tend to be suppressed in the presence of
stripes\cite {wan01, eme00, pre01, lor02b}.

\begin {figure}
\includegraphics [width=5cm,angle=0,clip=true]{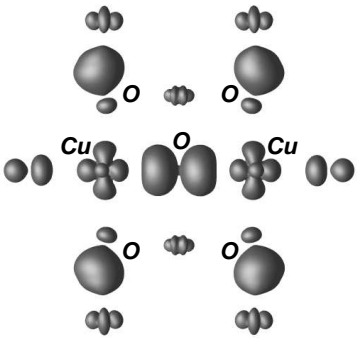}
\caption {Wannier function of the active band.}
\label {wf}
\end {figure}

A popular model to study the stability and symmetry of stripes in cuprates
is the single-band Hubbard model and its strong coupling version, the $t-J$
model\cite {whi98prl80, whi98prl81, whi03}. These models are usually
derived by considering the low-energy many-body states of a single hole
added to the insulator. Zhang and Rice (ZR)\cite {zha88} pointed out that
the hole will occupy a particular linear superposition of the four $p$ oxygen 
orbitals pointing toward a Cu that has the same symmetry as the 
Cu $d_{x^2-y^2}$ orbital. The hole in this ``ZR-O orbital'' and the hole
on the Cu form a ZR singlet. 
The other (orthogonal) combinations of $p$ oxygen orbitals 
are projected out.  By doing so the
ZR-O orbital is not allowed to polarize. Such approximation is justified in
the dilute limit however when one considers the stability of dense phases
special care must be taken on defining low-energy models. In particular for
CDW phases, polarization involving high energy states can significantly
affect the stability and symmetry\cite {bri95, mei95}. Indeed this problem
involves the short range (high energy) physics of the system and the above
procedure may be inadequate.

To illustrate the problem we have computed the Wannier function
(WF) of the active band (Fig.~\ref {wf}). The WF is centered on
the O forming the stripe core but extends considerably to the
surrounding oxygens. It is clear that the Cu-centered ZR-O
orbitals are strongly polarized and so the usual Hubbard or $t-J$
mapping becomes unreliable. This is in contrast with what was
found for one hole added to the AF in LDA+$U$ or inhomogeneous HF
where the solution did involve a  ZR-O orbital\cite {ani92,
yon92}. We conclude that the stability of stripes can not be
reliably studied within the $t-J$ or single-band Hubbard model.
(Of course the possibility of stripes in those models is a problem
of interests in itself but does not have direct experimental
relevance). In particular configurations centered in the O will be
unfavored since involve a strong O polarization and single band
models neglect the relaxation energy involved. If one refrains
from studding the stability of stripes one can still assume
stripes and use an effective model to describe the low-energy
physics. 

Coulomb effects (including the polarization effects mention above)
produce a splitting of the Madelung potential for a hole in a 
Cu in the metallic region and a Cu in the AF region of 0.2eV
favoring the charge segregation.  In LDA this splitting is
static but a truly many-body treatment should take into account the 
dynamical effects of the polarization.

Other parameters 
are similar to the ones in the dilute limit\cite{hyb90, ani02}. i.e.
The metallic regions can be described by a two-leg Hubbard model
with a hopping along the legs and along the rungs 
$t_{\parallel}= t_{\perp}=0.52$~eV and a hopping along the 
diagonals $t^\prime=-0.08$~eV. By treating this model in
mean-field with a  filling of 3/4 electrons and fitting the 
gap to the upper bands we obtain $U=4.5$~eV 
for the Hubbard on-site interaction, similar to
the value in the dilute case ($ \sim 5$~eV)\cite {hyb90}.

The active band is formed by the odd  combination of the leg
orbitals. The associated Wannier functions are quite different for
odd and even combinations, in particular only the odd combination
can have admixture with the central $p_x$ orbital as shown in
Fig.~\ref {wf} which shows again that in terms of ZR-O orbitals
large relaxations are involved not taken into account in the
standard Hubbard model which weights all O's around a Cu on an
equal manner.

The two bands immediately below the active one are formed by the
orbitals of the AF ladders and are completely filled in agreement 
with the assumption that the environment is an 
insulator\cite{eme97,kiv98,gra99,liu01}. Therefore the AF regions 
 can be viewed as AF ladders. We computed the magnetic
interaction parameters according to prescription of Ref.~\cite {lie95}.
The interaction  along the rungs, $J_{\perp}=146$~meV, results to
be enhanced respect to the value obtained by doing the same
computation in the stoichiometric system $J=109$~meV\cite {ani02}
whereas the interaction along the legs, $J_{\parallel}=78$~meV, is
slightly reduced. If one consider the leg isolated from the 1DEG
this would lead to a large spin gap of order $\sim 95$~meV\cite{dag96}.

Coming back to the metallic regions, it is believed that transverse hopping
will drive an array of 1DEG to a two dimensional Fermi liquid or an ordered
state\cite {cas92}. However according to Ref.~\cite {eme97, kiv98, eme00,
gra99} when one couples the magnetic system and the 1DEG a spin gap will be
induced in the 1DEG. This effect will block the effective transverse hopping
but not the pair hopping\cite {eme97, kiv98, eme00}. The system can become a
sliding Luttinger phase and eventually a superconductor. Other related
scenarios include a boson-fermion model where the bosons are composed
fermions coupled antiferromagnetically in the insulating regions\cite
{net01} and a phase with  bond order but with two-dimensional fermions\cite
{voj99, sac03}. The large values of the AF interaction in the insulating
regions favor all these scenarios. However to make more progress an
estimate of the hybridization perpendicular to the stripes is essential.

\begin {figure}
\includegraphics [width=7cm,clip=true]{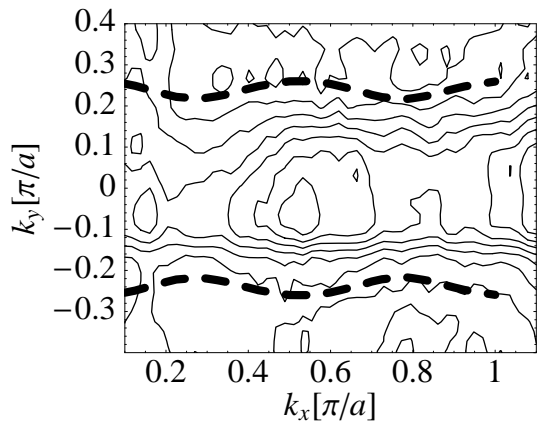}
\caption {Angle resolved photoemission spectral weight integrated
within 500~meV of the Fermi level, as a function of $k_x$ and
$k_y$\cite{zho99} together with calculated
Fermi surface for 2D CuO$_2$ plane with stripes oriented along 
$y$ (dashed line)\cite {note}.
Notice that the experimental Fermi surface has contributions due to stripes 
oriented along $y$ and along $x$ the latter producing the 
 vertical structures at  $k_x\sim 0.2 \pi/a$.
}
\label {fig:fs}
\end {figure}

The crossing of the active band in the $\Gamma$-B direction in
Fig.~\ref {bands} does not occur exactly at the middle point (as would be
the case for a perfectly half-filled one-dimensional band) but is
shifted to the B point. This shift depends on the momentum
perpendicular to the stripe and is precisely due to interstripe
hopping. i.e holes from the core of one stripe can tunnel to the
next stripe through the AF insulating like region. As a result the
Fermi surface is not flat but it warps as shown in Fig.~\ref
{fig:fs}. Interestingly this explains oscillations exhibit on the
experimental Fermi surface in the presence of stripes\cite
{zho99}. In Fig.~\ref {fig:fs} we show the experimental and the
theoretical data. The amplitude and the periodicity of the
wrapping is in good agreement with experiment. The large Fermi
surface encounter at larger doping has been shown to be in
agreement with LDA\cite{dam03}. The present result shows that in
underdoped materials LDA based methods can also predict the Fermi
surface including fine details.

The Fermi surface wrapping can be modeled by adding an effective
hybridization $t_{\perp,{\rm eff}}$ from core to core to the effective
two-leg Hubbard model mention above. This gives a contribution to the
single electron dispersion of the form $-2 t_{\perp,{\rm eff}}\cos(4a kx)$
with $t_{\perp,{\rm eff}}=0.015$~eV. Notice that this effect does not break
the approximate particle-hole symmetry of the system relevant for transport
experiments. The effective hybridization at the Fermi surface will be of
course renormalized by many-body effects of the kind considered in
Ref.~\cite {eme97, kiv98, eme00} which will tend to suppress it or hopping
with spin-wave fluctuations\cite {svr88} which will tend to enhance it.
The fact that the wrapping observed is close to the experimental one shows
that our estimate is not far from the final value although one should take
into account that the experiment is not done strictly at zero energy but
integrating in a finite window. More theoretical work is needed to 
decide the nature of the ground state with this value of the
transverse hopping. 

To conclude we have presented an LDA+$U$ study of vertical stripe in
cuprates. Stripes result to be bond-centered and quite wide in contrast
with the traditional stripe picture\cite {tra97}. The electronic structure
shows approximate particle-hole symmetry around the Fermi level in good
agreement with the picture deduced from transport experiments\cite {nod99,
wan01}.  The system can be pictured as alternating metallic Hubbard and
insulating Heisenberg ladders. O polarization effects are
important and, at the static level, they 
contribute to a large Madelung potential
splitting between insulating and metallic regions.  The dynamical
consequences should be further explored.   A large
exchange interaction in the insulating regions favor mechanisms for
superconductivity based on preformed AF singlets which can induce
superconductivity on the metallic regions. A non-negligibly hopping
transverse from stripe to stripe exist which produces observable effects
and is at the core of the low-energy QFT theories. This produces wrapping of
the Fermi surface which is in agreement with experimental observation. We
believe numerical studies of simplified models taking into account these
ingredients should shed light on the problem of superconductivity.


\begin {acknowledgments}
We thank O.K. Andersen, C. Castellani, C. Di Castro, G. Seibold
and M. Grilli for helpful discussions and Z.-X. Shen and
X.~J. Zhou for providing us the experimental data shown in Fig.~\ref{fig:fs}.
 This work has been
supported by RFFI grant 01-02-17063.
\end {acknowledgments}

\end {document}